\documentclass[a4paper,twoside]{article}
\newcommand{\affil}[1]{$^{\rm #1}$}
\usepackage[authoryear]{natbib}
\bibpunct{(}{)}{;}{a}{}{,}
\usepackage{graphicx}
\date{} 
\title{\large\bf\flushleft The Nature of V1464 Aql: \\
A New Ellipsoidal Variable with a $\delta$ Scuti Component}
\author{\parbox{\textwidth}{\flushleft
\vspace{-0.5cm}
{\it DAL, H.A. \affil{1},\affil{2}, S\.{I}PAH\.{I}, E. \affil{1} } \\
\vspace{0.4cm}
{\small \affil{1}\,Department of Astronomy and Space Sciences, University of Ege, Bornova, 35100 ~\.{I}zmir, Turkey}\\
{\small \affil{2}\,Corresponding Author, Email: ali.dal@ege.edu.tr}}}
\begin{document}
\twocolumn[
\begin{changemargin}{.8cm}{.5cm}
\begin{minipage}{.9\textwidth}
\vspace{-1cm}
\maketitle
%
%
\small{\bf Abstract:} Taking into account the results obtained from the models and analyses of the BVRI light curves, we discuss the nature of V1464 Aql. The analyses indicated that the mass ratio of the system is $q=0.71\pm0.02$, while the inclination of the system ($i$) is 38$^{\circ}$.45$\pm$0$^{\circ}$.22. Taking the primary component's temperature as 7420$\pm$192 K, we found that the temperature of the secondary is 6232$\pm$161 K. The mass of the primary component was found to be 1.74$\pm$0.05 $M_{\odot}$, while it is 1.23$\pm$0.01 $M_{\odot}$ for the secondary. The primary component's radius was found to be 2.10$\pm$0.05 $R_{\odot}$, while it was found as 1.80$\pm$0.01 $R_{\odot}$ for the secondary. Revealing that the system should not exhibit any eclipses, we demonstrated that the main variation with large amplitude should be caused due to the ellipsoidal effect. Indeed, the Fourier analysis also supported the result. For the first time in the literature, we revealed that the primary component is a $\delta$ Scuti star. The period of pulsation was found to be 58.482$\pm$0.002, 58.482$\pm$0.001, 60.966$\pm$0.002, 60.964$\pm$0.003 minutes in BVRI bands, respectively. We plotted V1464 Aql in the plane of $log(P_{orb})$-$log(P_{pulse})$. Using more than 160 binaries, whose one or both components are pulsating, we derived a new linear fit in the plane of $log(P_{orb})$-$log(P_{pulse})$ for each type binary. Using the linear fit of each group, we obtained new calibrations between $log(P_{orb})$ and $log(P_{pulse})$ for different type pulsating stars. In addition, a calibration has been obtained for the first time for the pulsating stars from the spectral types O and B. V1464 Aql seems to be located near the other ellipsoidal and close binaries. Thus, we listed V1464 Aql as a new candidate for the ellipsoidal variables with a $\delta$ Scuti component.\\


\medskip
\medskip
\end{minipage}
\end{changemargin}
]
\small

\section{Introduction}

The pulsation of a star, as well as being a component of eclipsing binary itself, is very important nature to understand its evolution. There are several type pulsating stars such as Cepheid-type pulsating stars, $\gamma$ Doradus-type pulsating stars, and $\delta$ Scuti-type pulsating stars and etc. All these types are in the Instability Strip on
the main-sequence in the Hertzsprung-Russell diagram. Investigating the frequencies of oscillations, asteroseismology known as stellar seismology tryies to identify the physical processes behind the pulsating nature and so the stellar interiors. Pulsating stars with their key roles are important objects to understand stellar evolution \citep{Cun07, Aer10}. However, in some cases, the pulsation features can not be enough to solve all the stellar interior and its nature. In this point, being a component of a binary can take an important role to get down the problems. Especially, analysis of a light curve together with a radial velocity gives us lots of physical parameters such as mass, radius, $log~g$, and ect., for the components of the eclipsing binaries \citep{Wil71, Wil90}. Considering the both pulsating and eclipsing behaviour, the physical natures of stars can be easily identifiable \citep{Lam06, Pig06}. Although there are many pulsating-single stars, a few of them are a component in a binary system, especially eclipsing system. As seen from the literature, the number of pulsating stars is over 635, while the incidence of eclipsing binary component all among them is $\sim$5.5$\%$ \citep{Kim10}. 

In this study, we introduce V1464 Aql as a new candidate for a binary with pulsating component. In the SIMBAD database, V1464 Aql was classified as a spectroscopic binary system from the spectral type of A2V. In the ASAS Database \citep{Poj97}, V1464 Aql's variability type was suggested as possibly RR Lyr (RRc), or an eclipsing binary, possibly contact (EC) or semi-detached system (ESD). The system was listed as a contact binary by \citet{Due97} and \citet{Pri06} and a spectroscopic binary by \citet{Pou04}. \citet{Due97} listed the period of V1464 Aql as 0$^{d}$.6978, while \citet{Ruc06} stated it as 0$^{d}$.697822. The detailed spectroscopic observations were made by \citet{Ruc06}. They obtained a single-lined radial velocity with amplitude of 30.62$\pm$2.35 $kms^{-1}$. The authors noted that the lines come from visible component are largely broadened. The value of $v \sin i$ was found to be 94$\pm$4 $kms^{-1}$. They also claim that the spectral type of the system should be F1-2 rather than early A.

In this study, we investigate the nature of this interesting system. For this aim, we observed system photometricaly, and obtained the multi-band light curves, in which the short-term intrinsic light variations are seen, for the first time. We analysed the light curves, and also tested the results with the method based on the Fourier transform described by \citet{Mor85} and \citet{Mor93}. Apart from light curve analyses, we also detected some large-amplitude pulsation with short period. We analysed the pulsating behaviour.

\section{Observations}

Observations were acquired with a thermoelectrically cooled ALTA U+42 2048$\times$2048 pixel CCD camera attached to a 40 cm - Schmidt - Cassegrains - type MEADE telescope at Ege University Observatory. The observations were made in BVRI bands during 11 nights in 03, 06, 07, 09, 16, 17 August 2011 and 04, 08, 15, 19, 30 September 2011. V1464 Aql was observed during all the night in each observations. Some basic parameters of the program stars are listed in Table 1. The names of the stars are listed in the first column, while J2000 coordinates are listed in the second column. The V magnitudes are in the third column, and B-V colours are listed in the last column.

Although the program and comparison stars are very close on the sky, using the calibration described by \citet{Har62}, the differential atmospheric extinction corrections were applied. The atmospheric extinction coefficients were obtained from observations of the comparison stars on each night. Heliocentric corrections were also applied to the times of the observations.

The mean averages of the standard deviations are 0$^{m}$.023, 0$^{m}$.011, 0$^{m}$.010, and 0$^{m}$.013 for observations acquired in the BVRI bands, respectively. To compute the standard deviations of observations, we used the standard deviations of the reduced differential magnitudes in the sense comparisons minus check stars for each night. There is no variation observed in the brightness of comparison stars.

The obtained BVRI data indicated that the system exhibits two main variations in each band. One of them has large amplitude with longer period, while the second one has relatively smaller amplitude with shorter period. We firstly investigated the larger variation. For this aim, we determine the most symmetric and deeper minimum. The light curve obtained in the observations on September 04, 2011 has such a minima. Using two different methods described by \citet{Kwe56} and \citet{Win67}, we computed a minimum time. Both methods gave the same minimum time with almost the same error. Then, using the method of the Discrete Fourier Transform (DFT) \citep{Sca82}, we analysed its period by the PERIOD04 software \citep{Len05}. The results obtained from DFT were tested by two other methods. One of them is CLEANest, which is another Fourier method \citep{Fos95}. The second method is the Phase Dispersion Minimization (PDM), which is a statistical method \citep{Ste78}. These methods confirmed the results obtained by DFT.

\begin{center}
\begin{equation}
JD~(Hel.)~=~24~55809.3467(1)~+~0^{d}.697822(3)~\times~E.
\end{equation}
\end{center}

All the found ephemeris of the system are given by Equation (1). The analyses reveal that the period of the larger variation is close to the ones given by \citet{Due97} and \citet{Ruc06}. All the observation were phased with using the ephemeris to prepare the light curve analyses.

\section{Light Curve Analyses}

\subsection{Eclipsing Binary Model}

In the literature, V1464 Aql was classified as a contact binary \citep{Due97, Pri06}. In addition, it was noted in the ASAS Database \citep{Poj97} that the system could also be an eclipsing binary, especially a contact binary. Considering this knowledge, we tried to model the light curves with the Wilson-Devinney Code \citep{Wil71, Wil90}. For this aim, we analysed the light curves obtained in the BVRI bands together with the available single-lined radial velocity curve simultaneously, using the PHOEBE V.0.31a software \citep{Prs05}, which is used in the version 2003 of the Wilson-Devinney Code \citep{Wil71, Wil90}. We tried to analyse the light curves with several different modes, such as the "overcontact binary not in thermal contact" and "double contact binary" modes. The initial analyses demonstrated that an astrophysical acceptable result can be obtained if the analysis is carried out in the "double contact binary" mode, while no acceptable results could be obtained in the other modes.

In the literature, the temperature is given in a range from 6843 K \citep{Amm06} to 8970 K \citep{Wri03} for V1464 Aql. We took JHK brightness of the system ($J=7^{m}.963\pm0^{m}.027$, $H=7^{m}.830\pm0^{m}.027$, $K=7^{m}.788\pm0^{m}.020$) from the NOMAD Catalogue \citep{Zac05}. Using these brightnesses, we derived de-reddened colours as a $(J-H)_{\circ}$=$0^{m}.110\pm0^{m}.038$ and $(H-K)_{\circ}$=$0^{m}.030\pm0^{m}.034$ for the system. Using the calibrations given by \citet{Tok00}, we derived the temperature of the primary component as 7420$\pm$192 K depending on these de-reddened colours. Taking into account the standard deviation of the brightness in JHK bands, we have computed the standard deviation of the temperature as 192 K. Thus, the temperature was found to be $7420\pm192$ K. Thus, the temperature of the primary component was fixed to 7420$\pm$192 K in the analyses, and the temperature of the secondary was taken as a free parameter. Considering the spectral type corresponding to this temperature, the albedos ($A_{1}$ and $A_{2}$) and the gravity darkening coefficients ($g_{1}$ and $g_{2}$) of the components were adopted for the stars with the convective envelopes \citep{Luc67, Ruc69}. The non-linear limb-darkening coefficients ($x_{1}$ and $x_{2}$) of the components were taken from \citet{Van93}. In the analyses, their dimensionless potentials ($\Omega_{1}$ and $\Omega_{2}$), the fractional luminosity ($L_{1}$) of the primary component, the inclination ($i$) of the system, the mass ratio of the system ($q$) and the semi-major axis ($a$) were taken as the adjustable free parameters.

Because of being single-lined, the available radial velocity curve does not give any mass ratio for the system. Therefore, considering general sum of weighted squared residuals ($\Sigma res^{2}$), we tried to find the mass ratio for the system. In analyses, the $\Sigma res^{2}$ values indicated that the photometric mass ratio of the components is $q=0.71\pm0.02$. According to this result, we assume that the mass ratio of the system is $q=0.71\pm0.02$. The temperature of the secondary component was found to be 6232 K and its error was found to be 27 K. However, the found error seems to be unreal in the sense of statistical. Taking into account the standard deviation of the brightness in JHK bands, we have computed the standard deviation of the temperature as 161 K. Finally, the temperature was found to be $6232\pm161$ K for the secondary component. All the parameters derived from the analyses are listed in Tables 2, while the synthetic light curves are shown in Figure 1. In addition, the radial velocity obtained by \citet{Ruc06} and the synthetic curve is shown in Figure 2. Finally, we also derived the 3D model of Roche geometry and the geometric configurations at four special phases 0.00, 0.25, 0.50 and 0.75 for the system, using the parameters obtained from the light curve analysis. The derived 3D model of Roche geometry and the geometric configurations are shown in Figure 3.

Although there is not any available double lined-radial velocity curve, we tried to estimate the absolute parameters of the components. According to \citet{Tok00}, the mass of the primary component must be 1.74$\pm$0.05 $M_{\odot}$ corresponding to its surface temperature. Considering possible mass ratio of the system, the mass of the secondary component was found to be 1.23$\pm$0.01 $M_{\odot}$.

Using Kepler's third law, we calculated possible the semi-major axis as a 4.76$\pm$0.03 $R_{\odot}$. Considering this estimated semi-major axis, the radius of the primary component was computed as 2.10$\pm$0.05 $R_{\odot}$, while it was found as 1.80$\pm$0.01 $R_{\odot}$ for the secondary component. Using the estimated radii and the obtained temperatures of the components, the luminosity of the primary component was estimated to be 11.97$\pm$0.31 $L_{\odot}$, and it was found as 4.41$\pm$0.02 $L_{\odot}$ for the secondary component. We plotted the distribution of the radii versus the masses and also the luminosity versus the temperature for some stars in Figure 4. The lines represent the ZAMS theoretical model developed for the stars with $Z=0.02$ by \citet{Gir00}, while dashed lines represent the TAMS theoretical model. According to the obtained results, the absolute parameters are generally an acceptable in the astrophysical sense.

\subsection{Elipsoidal Binary Model}

As seen from the Roche geometry shown in Figure 3, and the results of the light curve analysis, the system does not exhibit any eclipses. However, the shape of the components should causes some light variation known as the ellipsoidal variation. The ellipsoidal variable stars are non-eclipsing binary stars and especially close binaries \citep{Mor85, Bee85}. In the case of the ellipsoidal variables, the inclination angle ($i$) of the binary is so small that the system does not exhibit any eclipses. The main variation is due to non-spherical shapes of the components. According to \citet{Mor85}, \citet{Bee85} and \citet{Mor93}, if the main variation is caused due to the ellipsoidal effect indeed, the light curves of the ellipsoidal variables can be modelled by the Fourier analysis given by Equation (2), and also one will expect that the $\cos (2 \theta)$term must be dominant among all other terms in the results of the Fourier analysis. This is the base of the modern methods used in recent studies, such as \citet{Fai11} and \citet{Fai12}.

\begin{center}
\begin{equation}
\label{eq:five}
L(\theta)= A_{0} ~ + ~ \sum_{\mbox{\scriptsize\ i=1}}^N ~ A_{i} ~ cos(i \theta) ~ + ~ \sum_{\mbox{\scriptsize\ i=1}}^N ~ B_{i} ~ sin(i \theta)
\end{equation}
\end{center}

All the BVRI light curves were modelled with the Fourier series. The derived Fourier models are shown in Figure 1, and the Fourier Coefficients are listed in Table 3. The $A_{i}$ coefficients listed in the table are the coefficients of the $\cos (i \theta)$ terms, while $B_{i}$ parameters are the coefficients of the $\sin (i \theta)$ terms given in Equation (2). In fact, the most dominant one is $\cos (2 \theta)$ term for each of the BVRI-bands. Thus, it is obvious that the main effect seen in the light variations is the ellipticity effect. However, if the most dominant one was $\cos (\theta)$ term, we would consider the other effects such as magnetic activity occurring on the surface of cool stars. There are many systems, which are similar to V1464 Aql, such 75 Pegasi and 42 Persei \citep{Mar90, Mar91}, and several other systems \citep{Fai11, Fai12}.

\section{The Pulsation}

Apart from the main variation, there is a shorter period oscillation. In order to understand the shorter period oscillation, we firstly obtained all the pre-whitened light curves. To obtain the pre-whitened data, we extracted synthetic light curves from all the observation in each band. In the second step, all the pre-whitened data were analysed with both the Discrete Fourier Transform (DFT, \citealt{Sca82}) and the Phase Dispersion Minimization (PDM), which is a statistical method \citep{Ste78}.

The periods found from the pre-whitened data are listed in Table 4. The normalized power-spectrums, which exhibit the quality of the period analysis, are shown in Figure 5. Using the light elements given with Equation (1) and the parameters found from the pre-whitened data, all the nightly light variations were modelled for each band. The modelled light variations are seen in Figure 6.

\citet{Soy06} listed some eclipsing binaries, whose primary and/or secondary components are pulsating. Using the physical parameters found in this study, the location of V1464 Aql's primary component is shown in Figure 7. In this figure, following \citet{Soy06}, the ZAMS (broad line) and TAMS (dashed line) were taken from \citet{Gir00}. The borders of the Instability Strip on the main-sequence were computed from \citet{Rol02}. We show that V1464 Aql's primary component is located among the eclipsing binaries, whose one or two component(s) are in the Instability Strip on the main-sequence. The primary component of V1464 Aql is a pulsating variable for the case at least. In the literature, there are also four ellipsoidal variables with a $\delta$ Scuti component. As it is listed in Table 5, these are XX Pyx \citep{Aer02}, HD 173977 \citep{Cha04}, HD 207651 \citep{Hen04} and HD 149420 \citep{Fek06}. In this case, V1464 Aql is fifth ellipsoidal variable binary with a $\delta$ Scuti component.

As it is known from \citet{Tur11, Lia12}, the close binaries with a pulsating component exhibit some calibrations in the plane of $log(P_{orb})$-$log(P_{pulse})$. For this aim, we took the data of all listed close binaries, whose components are from A or F spectral types, from \citet{Tur11, Lia12}. We also took the data of listed binaries, whose components are generally from O or B spectral types, from \citet{Aer04}. In addition, we plotted all the Long Secondary Period Variables (hereafter LSPVs) listed by \citet{Kis99, Oli03} in the same plane. In the case of the binaries, whose components are from A and F spectral types, the pulsating components are $\delta$ Scuti stars, while the components are the other pulsating stars excepted $\delta$ Scuti star for binaries, whose components are generally from the spectral types O or B. Although the LSPVs are also debated subject in the literature, there is a relation between their periods \citep{Kis99, Oli03}. In Figure 8, we plotted all the binaries in the plane of $log(P_{orb})$-$log(P_{pulse})$. All the components plotted in this figure are pulsating stars. Moreover, we plotted the pulsating components of all known ellipsoidal variables, whose one component is a $\delta$ Scuti star.

\begin{center}
\begin{equation}
log(P_{pulse})~=~-0.040(0.070)~\times~log(P_{orb})~-~0.337(0.025)
\end{equation}
\end{center}

\begin{center}
\begin{equation}
log(P_{pulse})~=~0.593(0.041)~\times~log(P_{orb})~-~1.545(0.021)
\end{equation}
\end{center}

\begin{center}
\begin{equation}
log(P_{pulse})~=~1.083(0.024)~\times~log(P_{orb})~-~1.449(0.059)
\end{equation}
\end{center}

Using GraphPad Prism V5.02 software \citep{Mot07}, we re-modelled the distribution of each group with the linear fit. Here, the standard deviations of each coefficient and each constant are given in the brackets near themselves. The derived linear fits are also plotted in Figure 8. To test whether the derived linear fits are statistically acceptable, we computed the probability value (hereafter p-value). The threshold level of significance (hereafter $\alpha$ value) was taken as 0.005 for the p-value, which allowed us to test whether the p-value are statistically acceptable or not \citep{Wal03}. The derived linear fit is given with Equation (3) for the binaries, whose components are from spectral types O and B. The p-value was found to be $p-value$ $<$ 0.0027. The linear fit of the A-F binaries is given with Equation (4), while it is given with Equation (5) for the ellipsoidal variables. The computed $p-values$ are $<$ 0.0001 for both groups. Considering the $\alpha$ value, the derived linear fits are statistically acceptable.

As seen from Figure 8, in the plane of $log(P_{orb})$-$log(P_{pulse})$ distribution, V1464 Aql locates near the close binaries and ellipsoidal variables from the spectral types A and F.

\section{Results and Discussion}
	
In the literature, there are several stars or systems, which exhibits combinations of a few different variations. Some of them are eclipsing binaries with a pulsating component, and some semi-regular variable with unknown short-term variations, or some pulsating ellipsoidal variables, ect \citep{Der06, Nie10, Fai11, Fai12}.

In several studies, V1464 Aql was classified as a spectroscopic binary system, RR Lyr (RRc), an eclipsing binary, possibly contact (EC) or semi-detached system (ESD) \citep{Due97, Pri06, Pou04, Ruc06}. We analysed the BVRI light curves. As seen from Table 2, the inclination of the system ($i$) is found to be 38$^{\circ}$.45$\pm$0$^{\circ}$.22. In the case of this ($i$) value, the system does not exhibit any eclipses. As shown in Figure 3, the 3D geometric configurations derived at four special phases for the system also demonstrated the non-eclipsing cases. However, these configurations indicate that the system is on the edge of eclipsing. Although the system may exhibit the grazing eclipses, these eclipses become unseen among the other small variability. As it is estimated in the literature, the 3D model of Roche geometry reveals that the components of the system are near the filling their Roche Lobes. According to these results, V1464 Aql should be a candidate of contact binary system, but non-eclipsing due to the inclination of the system ($i$). The Fourier analysis also supported this result. The coefficients of the $\cos (2 \theta)$ term ($A_{2}$) were found to be 0.0492$\pm$0.0008, 0.0372$\pm$0.0007, 0.0355$\pm$0.0006 and 0.0327$\pm$0.0005 for the BVRI bands, respectively. According to these values, the main effect in the light variation of the system is come from the shape of the components, known as ellipsoidal effect.

The calibration given by \citet{Tok00}, the mass of the primary component should be 1.74$\pm$0.05 $M_{\odot}$ corresponding to its surface temperature. Under this assumption, the light curve analysis reveals that the temperature of the secondary component was found to be 6232$\pm$161 K. The light curve analysis indicate that the mass ratio of the system should be $q=0.71\pm0.02$. In this point it should be noted that the standard deviation of the primary component temperature and absolute parameters were computed depending on the standard deviations given for the colours in the NOMAD Catalogue \citep{Zac05}. However, the error of temperature was found from the light curve analysis for the secondary component. Considering the standard deviations of the other parameters, it seem a bit lower than it should be. The places of the components plotted in Figure 4 demonstrated that the results found from the analysis are acceptable in the astrophysical sense. On the other hand, the secondary component is seen more closer to the TAMS than the primary in Figure 4. The stars seem not to be coeval to each other. However, it is well known that there are several binaries such as YY\,CrB \citep{Ess10}, BS\,Cas \citep{Yan08}, VZ\,Tri \citep{Yan10}. All these samples exhibits the same behaviour. Their common properties are that all of them are contact binaries with a large the period variation due to the large mass transfer. Indeed, if V1464 Aql is an analogue of these systems, this makes V1464 Aql very important for the future studies.

The fractional radii were found to be $r_{1}=0.441\pm0.002$ for the primary component and $r_{2}=0.378\pm0.002$ for the secondary one. In this case, the sum of fractional radii was computed as $r_{1}+r_{2}\simeq0.80$. Thus, V1464 Aql seems to be in agreement with \citet{Kop56}'s criteria for overcontact systems. The period analysis indicates that the orbital period as $0^{d}.697822$. In addition, the temperature of the primary component is 7420$\pm$192 K, while the secondary one is 6232$\pm$161 K. Although some contact binaries have components with some different surface temperature, they generally have the same surface temperature. Here, the primary component of V1464 Aql is hotter than the secondary one. Considering some characteristics of the system such as the short orbital period, small mass ratio, hotter primary component and ect., V1464 Aql seems to be in agreement with those of A-type W UMa binaries \citep{Ber05, Ruc85}. The period analyses reveal that the period of the short-term variation was found to be 58.482$\pm$0.002, 58.482$\pm$0.001, 60.966$\pm$0.002, 60.964$\pm$0.003 minutes in BVRI bands, respectively. However, it must be noted that we did not find any secondary frequency for the short-term variation. The period differences between each band should be caused by the different sensitivity of each band. The sensitivity is decreasing from B band to I band, because the amplitude of the pulsation is decreasing from B band to I band. As it is seen from the standard deviations given in Table 4 and also from the light curves shown in Figure 6, the scattering in the light curves is increasing from B band to I band. The period analysing methods we used are depend on the statistical method. The analyses gave the best period statistically for the pulsation in each band. In this case, the most reliable periods are ones found from B and V bands.

Figure 5 indicates that the short-term variation dominately exhibits itself in shorter wavelength rather than longer ones. In fact, this is also seen clearly in Figure 6. The amplitude of the short-term variation get larger from the I-band to the B-band. Moreover, the primary component is located in the Instability Strip on the main-sequence, as seen in Figure 7. Considering the mass of the primary component and also the period of this short-term variation, this second variation should be caused due to the pulsation of the primary component. According to \citet{Aer10}, the periods of $\delta$ Scuti stars are in the range 18 min to 8 hr. Moreover, the mass of them is in the range 1.50 to 2.50 $M_{\odot}$. The mass value of the component and the period of the short-term variation are agreement with these values. In this case, this variation should be caused by the pulsation of a $\delta$ Scuti star.

As it is seen from Figure 1, the light curves seem to be scattered. There are also a few light curves in the ASAS Database \citep{Poj97}. They are also very scattered. However, the models of pulsation shown in Figure 6 demonstrates that the scattering in the light curves is caused due to the pulsation. 

In the literature, there are four systems, whose light curves have the same properties. All of them are also ellipsoidal variables, which have a $\delta$ Scuti component. \citet{Aer02} found more than 10 frequencies due to the $\delta$ Scuti component in XX Pyx, whose orbital period is 1$^{d}$.15. HD 207651 is a triple system. \citet{Hen04} demonstrated that its component A is an ellipsoidal binary with a $\delta$ Scuti component. They found a few pulsation frequencies, and the more possible one is 0$^{d}$.7354. \citet{Cha04} found $\sim$0$^{d}$.1169 for the pulsation in the case of HD 173977. According to the light curve analysis given by \citet{Fek06}, HD 149420 should be an eclipsing binary. However, it is likely to be an ellipsoidal variable. They found the pulsating periods of $\delta$ Scuti component as 0$^{d}$.076082 and 0$^{d}$.059256. In the case of V1464 Aql, we found pulsating period $\sim$0$^{d}$.040641.

On the other hand, in the literature, classifying of V1464 Aql is a debated subject \citep{Poj97, Due97, Pri06, Pou04}. Because of this, we plotted V1464 Aql among the different type binaries and variables in the plane of $log(P_{orb})$-$log(P_{pulse})$ shown in Figure 8. As it is seen in the figure, indeed, V1464 Aql locates near the close and the ellipsoidal binaries, whose components are from the spectral types A and F. This case also supports that the effective temperature found for the primary component of V1464 Aql is generally an acceptable in the astrophysical sense. In addition, it this point, we also derived some new calibrations for different type binaries. Both \citet{Tur11} and \citet{Lia12} have already derived the calibration for the close binaries. Especially, \citet{Lia12} derived the calibrations separately for the detached binaries, semi-detached binaries. As given with Equation (3), combining all the close binaries listed by both \citet{Tur11} and \citet{Lia12}, we derived it together for all the close binary from the spectral types A and F. Thus, we obtained more reliable calibrations, using more larger data set. In addition, in the literature, the calibration has not been derived for binary from the spectral types O and B. Although one of their components is pulsating in these binaries, they are generally different class instead of the $\delta$ Scuti stars. As given with Equation (4), we tried to derive a similar calibration for these stars for the first time in the literature. A similar calibration was derived for the ellipsoidal binaries, as given with Equation (5).

\citet{Ruc06} noted that there are largely broadened lines come from the visible component. They also found that the $v \sin i$ value of the component is 94$\pm$4 $kms^{-1}$. Although this rapid rotation can cause the broadening in the lines, the pulsation can also cause it. Moreover, the surface temperature of the primary component is 7420$\pm$192 K, while it is 6232$\pm$161 K for the secondary. In this case, it is expected that we see double lines in the spectrum, though the inclination of the system ($i$) is 38$^{\circ}$.45. However, \citet{Ruc06} stated that V1464 Aql is a single-lined system, and the amplitude of the radial velocity is 30.62$\pm$2.35 $kms^{-1}$. The given amplitude of the radial velocity can easily confuse someone's mind, because it is very large enough. In this case, we also expected to see the lines of the secondary component, and thus, the radial velocity of the secondary component. On the other hand, considering both the spectral and the photometrical results together, rapid rotation and pulsation nature of the primary component can cause some effects on its spectrum in the case of lower inclination ($i$) of 38$^{\circ}$.45. The broadening of the lines should make the lines come from the secondary component disappear.

Consequently, V1464 Aql is a new candidate for the ellipsoidal system with a pulsating component. The new studies will reveal its pulsating nature very well, thus they will reveal the absolute nature of the system.

\section*{Acknowledgments} The author acknowledges the generous observing time awarded to the Ege University Observatory. We also thank the referee for useful comments that have contributed to the improvement of the paper.

\clearpage

\begin{figure*}
\hspace{5.0 cm}
\includegraphics[width=20cm]{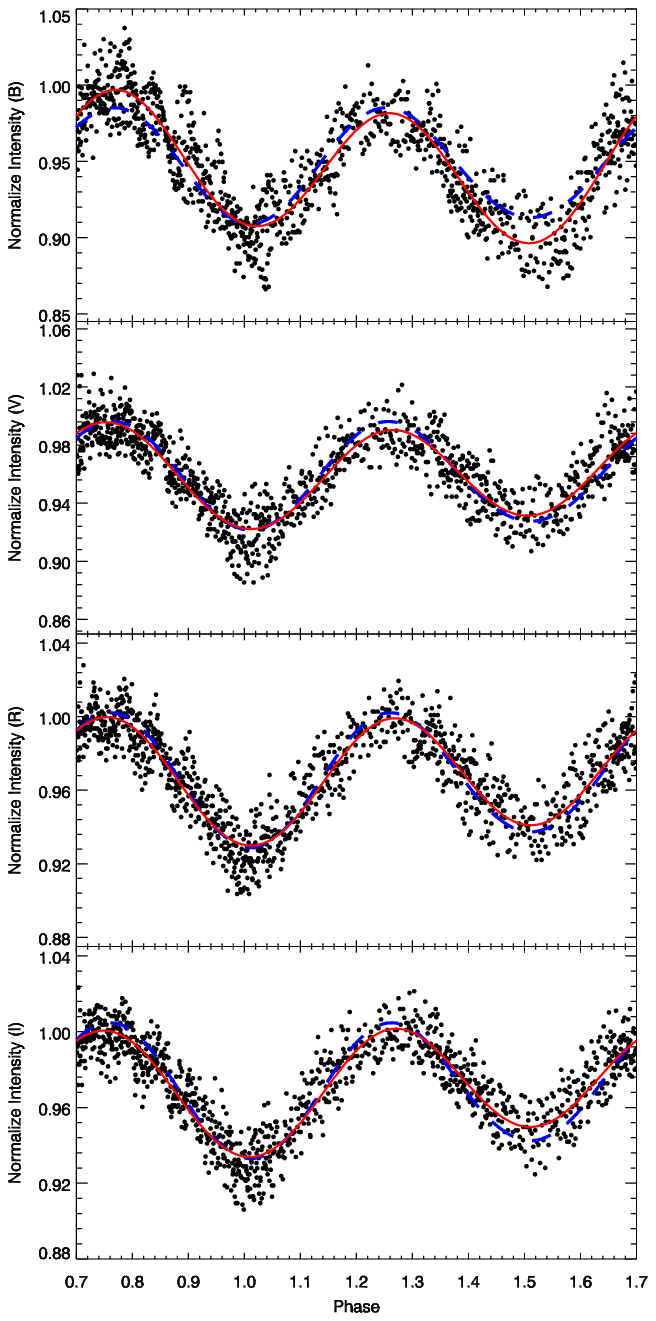}
\vspace{0.5 cm}
\caption{The observed BVRI light curves and the synthetic curves obtained from the light curve analysis and Fourier models. In the figure, the filled circles represent the observations, while the dashed (blue) lines represent the synthetic curves obtained from the light curve analysis and the (red) line represent the derived Fourier fits.}
\label{Fig. 1.}
\end{figure*}

\begin{figure*}
\hspace{2.75 cm}
\includegraphics[width=14cm]{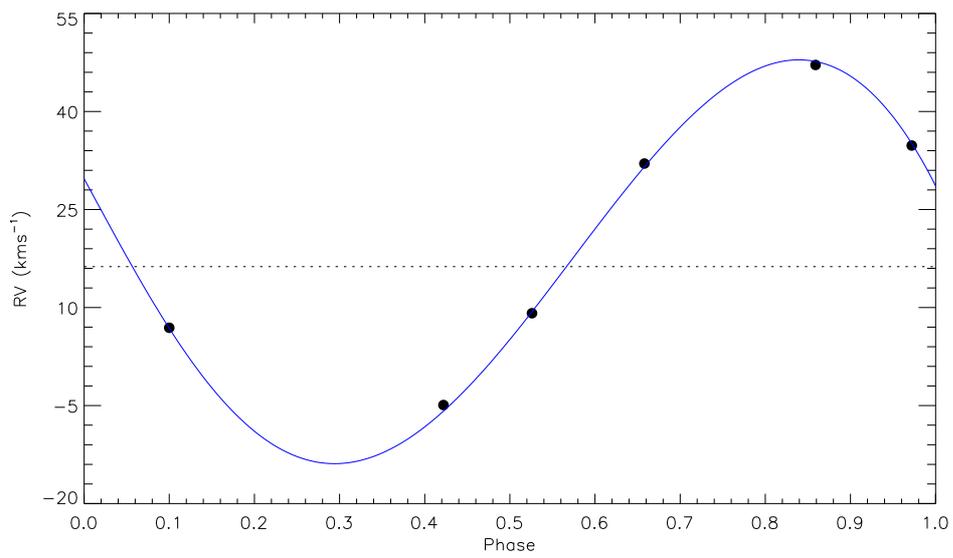}
\vspace{0.5 cm}
\caption{The radial velocity (filled circles) obtained by \citet{Ruc06} and the synthetic curve (blue line) derived by using the parameters found from the light curve analysis. In the figure, the dotted line represents the center-of-mass velocity.}
\label{Fig. 2.}
\end{figure*}

\begin{figure*}
\hspace{5.0 cm}
\includegraphics[width=6.0cm]{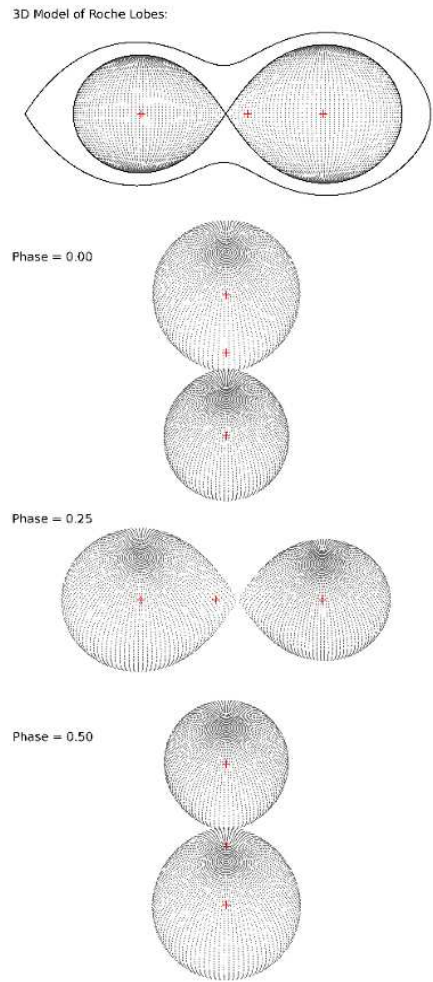}
\vspace{0.5 cm}
\caption{The 3D model of Roche geometry and the geometric configurations at four special phases 0.00, 0.25, 0.50 and 0.75, illustrated for V1464 Aql, using the parameters obtained from the light curve analysis.}
\label{Fig. 3.}
\end{figure*}

\begin{figure*}
\hspace{4.5 cm}
\includegraphics[width=17cm]{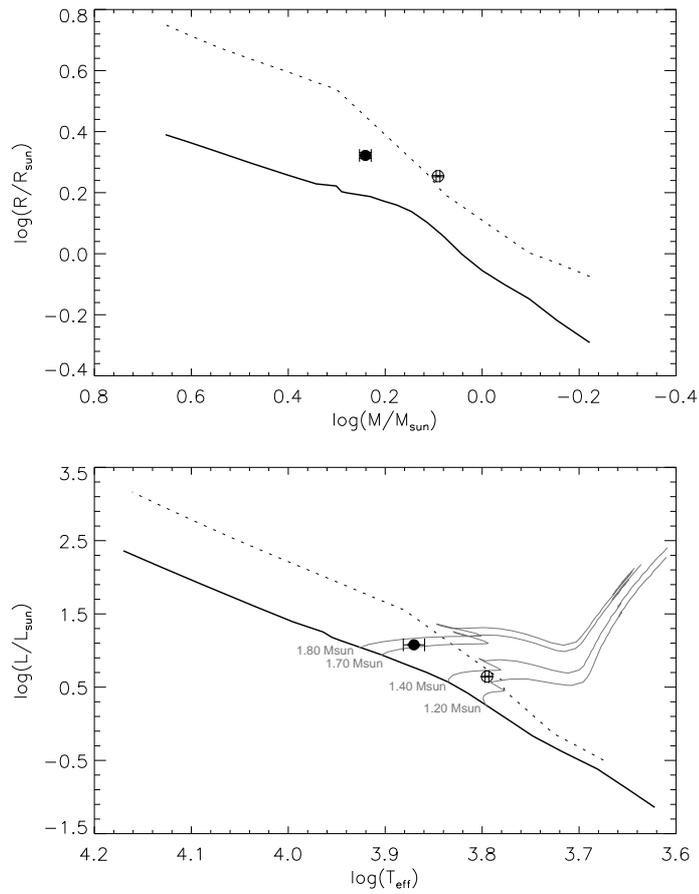}
\vspace{0.7 cm}
\caption{The places of the components of V1464 Aql in the Mass-Radius (upper panel) and Temperature-Luminosity (bottom panel) distributions. In the figure, the lines represent the ZAMS theoretical model developed by Girardi et al. (2000), while the dashed lines represent the theoretical TAMS model. The grey lines represent the theoretical evolutional tracks. The filled circles represent the primary component, while the open circles represent the secondary component.}
\label{Fig. 4.}
\end{figure*}

\begin{figure*}
\hspace{1.0 cm}
\includegraphics[width=15cm]{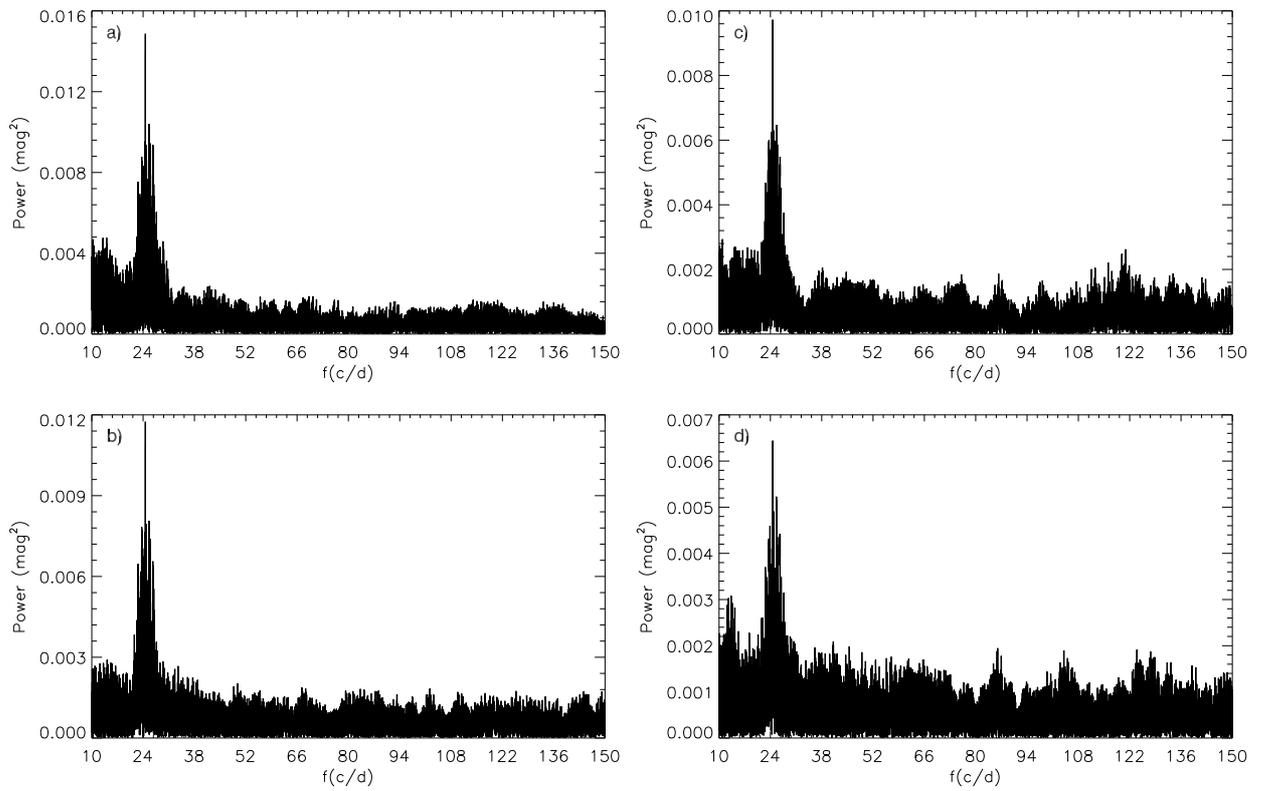}
\vspace{0.7 cm}
\caption{V1464 Aql's normalized power-spectrums obtained from the period analyses of the pre-whitened data with DPT method in each band. The power-spectrum obtained from BVRI-bands are given in the panels a, b, c, d, respectively.}
\label{Fig. 5.}
\end{figure*}

\begin{figure*}
\hspace{0.4 cm}
\includegraphics[width=23cm]{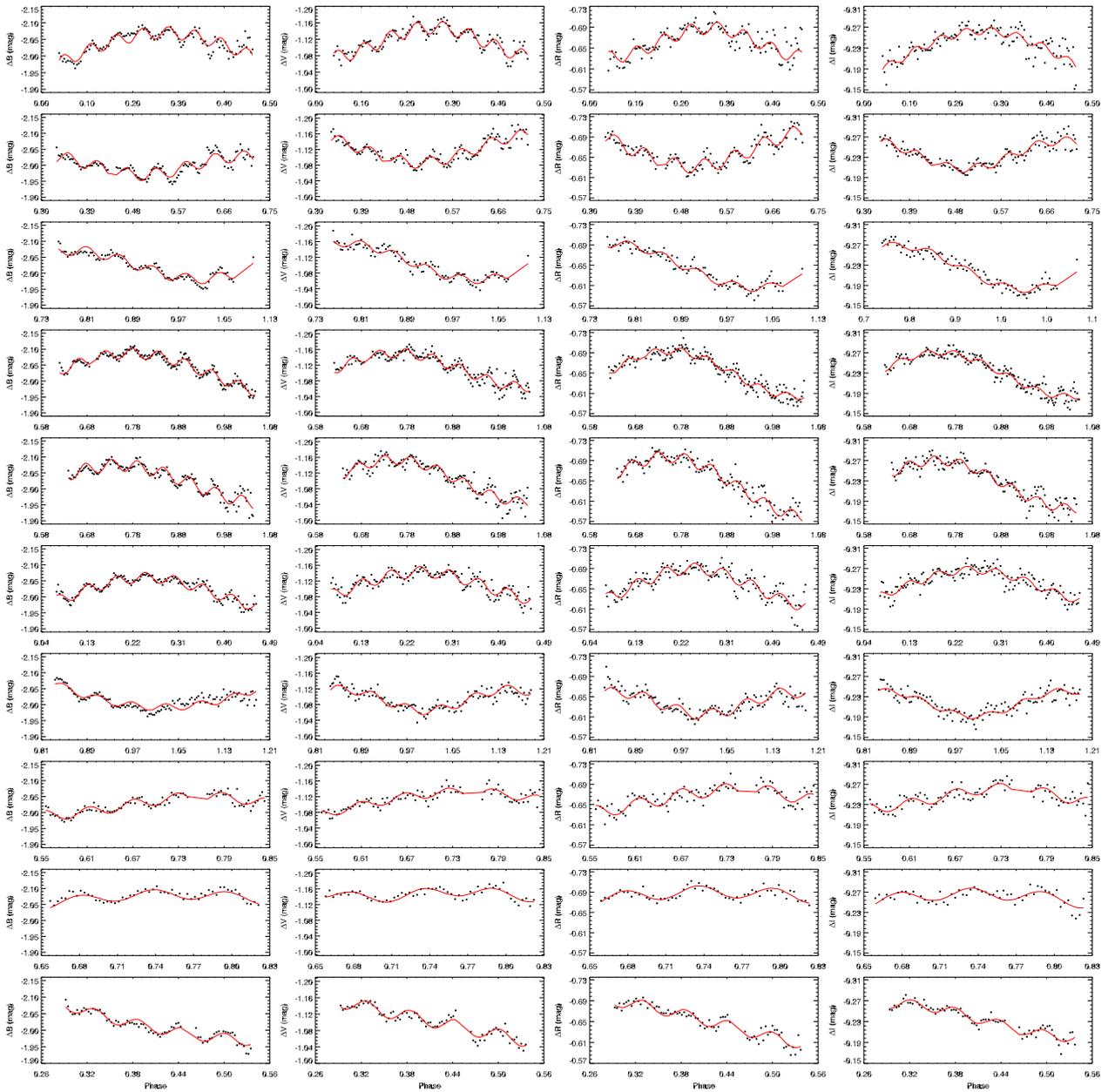}
\vspace{0.1 cm}
\caption{All the models of nightly light variations.}
\label{Fig. 6.}
\end{figure*}

\begin{figure*}
\hspace{3.0 cm}
\includegraphics[width=15cm]{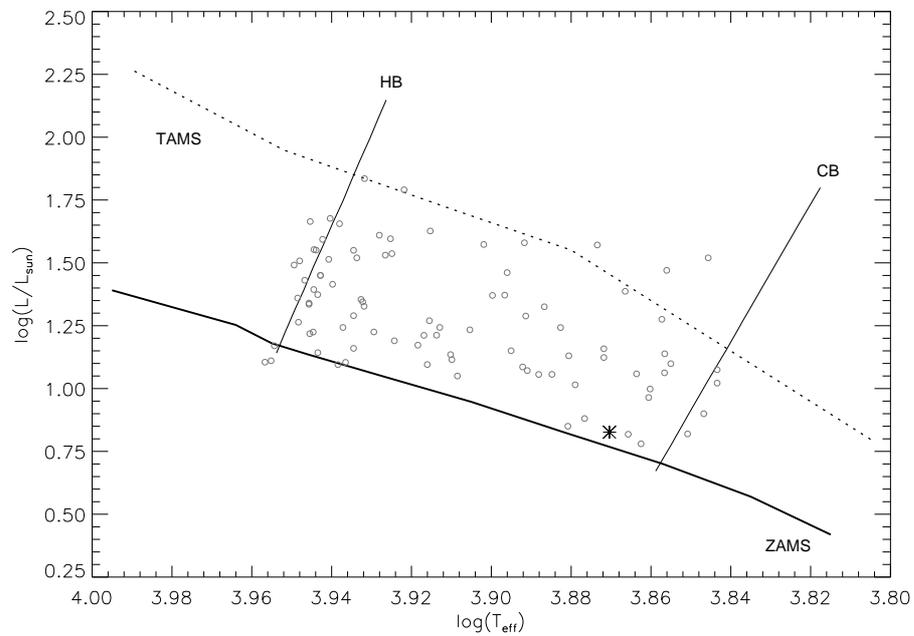}
\vspace{0.7 cm}
\caption{The location of V1464 Aql's primary component in the Instability Strip on the main-sequence. The symbol of asterisk represents the primary component of V1464 Aql, while the dim circles represent some semi- and un-detached binaries taken from Soydugan et al. (2006) and and references therein. In the figure, the ZAMS and TAMS were taken from Girardi et al. (2000), while the borders of the Instability Strip were computed from Rolland et al. (2002).}
\label{Fig. 7.}
\end{figure*}

\begin{figure*}
\hspace{3.0 cm}
\includegraphics[width=15cm]{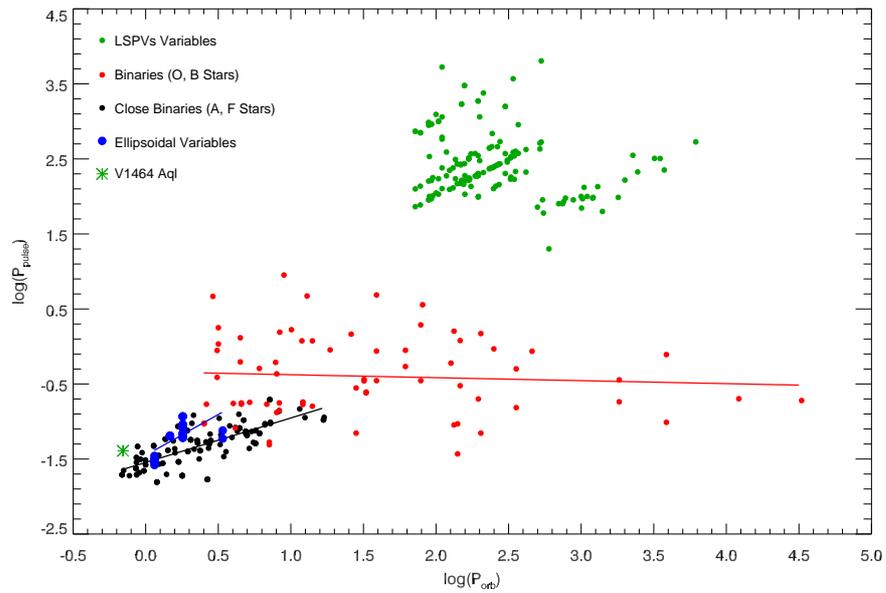}
\vspace{0.7 cm}
\caption{The location of V1464 Aql in the plane of $log(P_{orb})$-$log(P_{pulse})$ among different type binary stars. In the figure, filled circles represent the observed values, while the lines represent the theoretical linear fit of each group.}
\label{Fig. 8.}
\end{figure*}

\clearpage

\begin{table*}
\begin{center}
\caption{Basic parameters for the observed stars. The coordinates were taken from the SIMBAD database, the brightness and colours were taken from the All-sky Compiled Catalogue of 2.5 million stars \citep{Kha09}.}
\vspace{0.2 cm}
\begin{tabular}{lccc}
\hline
\hline
Star	&	$\alpha$ / $\delta$ (J2000)	&	V (mag)	&	B-V (mag)	\\
\hline							
V1464 Aql	&	19$^{h}$ 50$^{m}$ 15$^{s}$.473 / -08$^{\circ}$ 36$^{\prime}$ 06$^{\prime\prime}$.26	&	8.661	&	0.276	\\
GSC 5725 2283 (Comparison)	&	19$^{h}$ 49$^{m}$ 59$^{s}$.290 / -08$^{\circ}$ 35$^{\prime}$ 05$^{\prime\prime}$.86	&	10.137	&	1.123	\\
GSC 5725 2387 (Check) &	19$^{h}$ 50$^{m}$ 15$^{s}$.473 / -08$^{\circ}$ 36$^{\prime}$ 06$^{\prime\prime}$.26	&	9.860	&	1.272	\\
\hline
\end{tabular}
\end{center}
\end{table*}

\begin{table*}
\begin{center}
\caption{The parameters of the components obtained from the light curve analysis.}
\vspace{0.2 cm}
\begin{tabular}{lr}
\hline
\hline
Parameter	&	Value	\\
\hline
Orbital Period ($P$) & 0$^{d}$.697822 \\
Mass Ratio ($M_{2}/M_{1}=q$)	&	0.71$\pm$0.02	\\
Inclination ($i$)	&	38$^\circ$.45$\pm$0$^\circ$.22	\\
Temperature ($T_{1}$) &	7420$\pm$192$^{*}$ K	\\
Temperature ($T_{2}$) &	6232$\pm$161	K \\
Dimensionless Potential ($\Omega_{1}$)	&	3.2592$\pm$0.0006	\\
Dimensionless Potential ($\Omega_{2}$)	&	3.2582$\pm$0.0006	\\
Fractional Luminosity (L$_{1}$/L$_{T}$, $B$)	&	0.673$\pm$0.049	\\
Fractional Luminosity (L$_{1}$/L$_{T}$, $V$)	&	0.646$\pm$0.042	\\
Fractional Luminosity (L$_{1}$/L$_{T}$, $R$)	&	0.623$\pm$0.037	\\
Fractional Luminosity (L$_{1}$/L$_{T}$, $I$)	&	0.605$\pm$0.031	\\
Gravity-darkening Coefficients ($g_{1}$, $g_{2}$)	&	1.00, 0.32	\\
Albedo ($A_{1}$, $A_{2}$)	&	1.00, 0.50	\\
Limb-darkening Coefficients ($x_{1,bol}$, $x_{2,bol}$)	&	0.657, 0.657	\\
Limb-darkening Coefficients ($x_{1,B}$, $x_{2,B}$)	&	0.808, 0.809	\\
Limb-darkening Coefficients ($x_{1,V}$, $x_{2,V}$)	&	0.703, 0.704	\\
Limb-darkening Coefficients ($x_{1,R}$, $x_{2,R}$)	&	0.592, 0.592	\\
Limb-darkening Coefficients ($x_{1,I}$, $x_{2,I}$)	&	0.489, 0.489	\\
Fractional Radius ($<r_{1}>$)	&	0.441$\pm$0.002	\\
Fractional Radius ($<r_{2}>$)	&	0.378$\pm$0.002	\\
Absolute Luminosity (L$_{1}$) &	11.97$\pm$0.31 $L_{\odot}$ \\
Absolute Luminosity (L$_{2}$) &	4.41$\pm$0.02 $L_{\odot}$ \\
Semi-Amplitude of Radial Velocity ($kms^{-1}$)	&	30.8132$^{+0.3705}_{-0.3615}$	\\
Systemic Velocity ($kms^{-1}$)	&	16.2560$^{+0.2434}_{-0.2469}$	\\
\hline
\end{tabular}
\end{center}
$^{*}$ The error is computed depending on the standard deviations of de-reddened colours. \\
\end{table*}

\begin{table*}
\begin{center}
\caption{The coefficients derived from the Fourier model.}
\vspace{0.2 cm}
\begin{tabular}{cccccc}
\hline
\hline
Filter	&	$A_{0}$	&	$A_{1}$	&	$A_{2}$	&	$B_{1}$	&	$B_{2}$	\\
\hline											
B	&	-2.0271$\pm$0.0006	&	-0.0075$\pm$0.0008	&	0.0492$\pm$0.0008	&	0.0078$\pm$0.0008	&	0.0091$\pm$0.0008	\\
V	&	-1.1129$\pm$0.0005	&	0.0052$\pm$0.0006	&	0.0372$\pm$0.0007	&	0.0032$\pm$0.0006	&	0.0041$\pm$0.0006	\\
R	&	-0.6554$\pm$0.0004	&	0.0064$\pm$0.0006	&	0.0355$\pm$0.0006	&	0.0008$\pm$0.0006	&	0.0052$\pm$0.0005	\\
I	&	-0.2349$\pm$0.0004	&	0.0092$\pm$0.0005	&	0.0327$\pm$0.0005	&	0.0000$\pm$0.0005	&	0.0041$\pm$0.0005	\\
\hline
\end{tabular}
\end{center}
\end{table*}

\begin{table*}
\begin{center}
\caption{The pulsation period found from period analyses in each filter.}
\vspace{0.2 cm}
\begin{tabular}{ccc}
\hline
\hline
Filter	&	Period (minute)	&	Amplitude (mag)	\\
\hline					
B	&	58.482$\pm$0.002	&	0.030$\pm$0.002	\\
V	&	58.482$\pm$0.001	&	0.024$\pm$0.003	\\
R	&	60.966$\pm$0.002	&	0.017$\pm$0.006	\\
I	&	60.964$\pm$0.003	&	0.011$\pm$0.006	\\
\hline
\end{tabular}
\end{center}
\end{table*}

\begin{table*}
\begin{center}
\caption{The known ellipsoidal binaries having a pulsating component in the literature.}
\vspace{0.2 cm}
\begin{tabular}{lccccccccc}
\hline
\hline					
Stars 	&	Spectral	&	V	&	 $P_{orb}$ 	&	 $P_{pulse}$ 	&	 $M_{1}$ 	&	 $M_{2}$ 	&	 $R_{1}$ 	&	 $R_{2}$ 	&	 Ref. \\
	&	Type$^{1}$	&	(mag)$^{1}$	&	 (day) 	&	 (day) 	&	 ($M_{\odot}$) 	&	 ($M_{\odot}$) 	&	 ($R_{\odot}$) 	&	 ($R_{\odot}$) 	&	 \\
\hline																		
XX Pyx 	&	A4	&	11.490	&	1.150000	&	0.02624	&	1.85	&	 - 	&	1.95	&	 - 	&	 $^{2}$ \\
HD207651 	&	A5	&	7.210	&	1.470800	&	0.73540	&	 - 	&	 - 	&	 - 	&	 - 	&	 $^{3}$ \\
HD173977 	&	F2	&	8.070	&	1.800745	&	0.06889	&	1.87	&	1.30	&	2.87	&	1.42	&	 $^{4}$ \\
HD149420 	&	A8	&	6.874	&	3.394306	&	0.07608	&	 - 	&	 - 	&	 - 	&	 - 	&	 $^{5}$ \\
\hline
\end{tabular}
\end{center}
$^{1}$ Taken from the SIMBAD data base \\
$^{2}$ \citet{Aer02} \\
$^{3}$ \citet{Hen04} \\
$^{4}$ \citet{Cha04} \\
$^{5}$ \citet{Fek06} \\
\end{table*}


\end{document}